\documentclass[aps,preprint,a4paper,showpacs,showkeys,superscriptaddress]{revtex4-1}

\usepackage{latexsym}
\usepackage{amsmath,amssymb}
\usepackage{graphicx}
\usepackage{subfigure}
\usepackage{xcolor}
\usepackage{cancel}

\usepackage{hyperref}     
\hypersetup{colorlinks,%
  citecolor=blue,%
  linkcolor=cyan,%
  pdftex}

\begin{document}


\title{Black hole complementarity
with the generalized uncertainty principle in Gravity's Rainbow}


\author{Yongwan Gim}%
\email[]{yongwan89@sogang.ac.kr}%
\affiliation{Department of Physics, Sogang University, Seoul, 04107,
  Republic of Korea}%
\affiliation{Research Institute for Basic Science, Sogang University,
  Seoul, 04107, Republic of Korea} %

\author{Hwajin Um}%
\email[]{um16@sogang.ac.kr}%
\affiliation{Department of Physics, Sogang University, Seoul, 04107,
  Republic of Korea}%

\author{Wontae Kim}%
\email[]{wtkim@sogang.ac.kr}%
\affiliation{Department of Physics, Sogang University, Seoul, 04107,
  Republic of Korea}%

\date{\today}

\begin{abstract}
When gravitation is combined with quantum theory,
the Heisenberg uncertainty principle could be extended to
the generalized uncertainty principle accompanying a minimal length.
To see how the generalized uncertainty principle works in the context of black hole complementarity,
we calculate the required energy to duplicate information for the Schwarzschild black hole.
It shows that the duplication of information is not allowed
 and black hole complementarity is still valid
even assuming the generalized uncertainty principle.
On the other hand, the generalized uncertainty principle with the minimal length
could lead to a modification of the conventional dispersion relation in light of Gravity's Rainbow,
 where the minimal length is also invariant as well as the speed of light.
Revisiting the gedanken experiment,
we show that the no-cloning theorem for black hole complementarity can be made valid
in the regime of Gravity's Rainbow on a certain combination of parameters.

\end{abstract}

%



\maketitle


\section{Introduction}
\label{sec:intro}

The discovery of Hawking radiation emitted from a black hole \cite{Hawking:1974rv, Hawking:1974sw}
would lead to the information loss paradox  \cite{Hawking:1976ra}; however,
it could be solved for a distant observer outside the horizon by assuming that
 the Hawking radiation carries the black hole information.
In this assumption, the local observer (Bob) outside the horizon gathers the information of the infalling matter state
through the Hawking radiation
after a certain time which amounts to at least the Page time \cite{Page:1993wv},
and then he jumps into the black hole.
If the infalling observer (Alice) who has the information of the infalling matter state
   sends the message with the information
 to him inside the black hole,
then he may have the duplicated information,
which is the violation of the no-cloning theorem
in quantum theory.
The so-called black hole complementarity has been proposed as a solution to this cloning problem
to reconcile general relativity and quantum mechanics \cite{Susskind:1993if, Susskind:1993mu, Stephens:1993an}.
According to black hole complementarity, the cloning problem never occurs
since the observer inside the horizon is not in the causal past of
any observer who measures the information through the Hawking radiation
outside the horizon   \cite{Susskind:1993if}.
The specific gedanken experiment \cite{Susskind:1993mu} on the Schwarzschild black hole
proves that the required energy to correlate the observations of both sides of the horizon
exceeds the mass of the black hole.
 In other words,
the information has to be encoded into the message with super-Planckian frequency
to duplicate it.
Thus,
it turns out that the no-cloning theorem for black hole complementarity
is safe for the Schwarzschild black hole.

On the other hand, it has been claimed that the notion of quantum theory may need a revision
if one attempts to combine gravitation and quantum theory.
One of the possibilities is the generalized uncertainty principle (GUP)
which is the extended version of the Heisenberg uncertainty principle
by adding a term of uncertainty in position due to the gravitational interaction
\cite{Maggiore:1993rv,  Maggiore:1993zu, Scardigli:1999jh, Adler:1999bu}.
It could be derived from not only
general considerations of quantum mechanics and gravity \cite{ Hossenfelder:2012jw}
but also  string theory \cite{Veneziano:1986zf, Gross:1987ar, Amati:1988tn, Konishi:1989wk},
which gives rise to a minimal length of the order of the Planck length, $(\Delta X)_{\rm min} \sim \sqrt{\alpha_{\textrm{\tiny GUP}}} L_P$.
Many efforts have been devoted to studying various aspects of the GUP \cite{Ali:2009zq, Bojowald:2011jd, Feng:2016tyt, Hammad:2015dka,   Pedram:2015jsa, Li:2016mwq, Masood:2016wma, Faizal:2017dlb}.
Especially, the role of GUP was discussed in the context of the information loss problem \cite{Itzhaki:1995tc}.
Subsequently, it was also shown that
while black hole complementarity on the Heisenberg uncertainty principle could be violated
if the black hole evaporated by emitting a sufficiently large number $N$ of species of massless scalar fields
as the Hawking radiation,
the GUP can prevent the violation of black hole complementarity
assuming that the GUP parameter is proportional to the number of fields,
 $\alpha_{\textrm{\tiny GUP}} \sim N$ \cite{Chen:2014bva}.
However, the large number of species makes the GUP parameter very large, so that
the GUP effect should be too significant.

Now, we note that the existence of the minimal length in the GUP
would necessarily lead to
important modifications such as the black hole temperature and the Stefan-Boltzmann law.
The temperature of the black hole could be modified by the GUP \cite{Adler:2001vs},
and so there have been many applications to black hole systems \cite{Custodio:2003jp, Myung:2006qr, Kim:2007hf, Banerjee:2010sd, Carr:2015nqa}.
Moreover, the Stefan-Boltzmann law should also be corrected in order
to describe the black hole evaporation consistently
in the regime of the GUP
along with the modification of the temperature \cite{AmelinoCamelia:2005ik, Nouicer:2007jg}.
So, one might wonder how
black hole complementarity can be made valid even for a single scalar field of $N=1$.

On the other hand, the system governed by the GUP does not allow the length scale below the minimal length, which means that the GUP is combined with the doubly special relativity of the extended version of Einstein's special relativity  \cite{AmelinoCamelia:2000ge, AmelinoCamelia:2000mn}, where there are two observer-independent scales such as the minimal length and the speed of light. The most common illustration was presented to keep the relativistic energy-momentum relation in the framework of the doubly special relativity \cite{Magueijo:2001cr},
which gives rise to the modified dispersion relation (MDR) \cite{Magueijo:2002am}.
In the framework of the doubly special relativity,
the modification of the measure of integration in momentum space should be considered  under the deformed symmetries,
otherwise
 the MDR is only valid in one reference frame, implying a breakdown of 
 the relativistic symmetries.
 The notion could be promoted to the curved spacetime, where  the energy of the test particle deforms the general spacetime of the background geometry, which is named  Gravity's Rainbow \cite{Magueijo:2002xx}.
For the MDR \cite{ Aloisio:2005qt,  Girelli:2006fw, Garattini:2011kp, Majhi:2013koa, Kiyota:2015dla,Rosati:2015pga, Barcaroli:2015xda, Barcaroli:2016yrl}
as well as Gravity's Rainbow
\cite{Galan:2005ju,  Ling:2006az, Galan:2006by,  Ling:2008sy,  Garattini:2011hy,  Garattini:2012ec, Amelino-Camelia:2013wha, Barrow:2013gia,  Awad:2013nxa,   Ali:2014xqa,  Ali:2014qra,  Chang:2014tca, Gim:2014ira,  Gim:2015zra, Gim:2015yxa,  Hendi:2015vta, Hendi:2016wwj, Hendi:2016hbe, EslamPanah:2017ugi,  Hendi:2017vgo}, there have been extensive studies in order for exploring various aspects for black holes and cosmology.
So, it seems to be natural to address the issues related to
black hole complementarity with the GUP in the context of Gravity's Rainbow.

In Sec.~\ref{sec:Sch}, in a self-contained manner we recapitulate the well-established gedanken experiment
to determine the required energy for the duplication of the infalling information
on the Schwarzschild black hole with the Heisenberg uncertainty relation
and the ordinary dispersion relation along the line of Ref. \cite{Susskind:1993mu}.
In Sec.~\ref{sec:GUPDR}, we calculate the required energy for cloning the information
for the Schwarzschild black hole by using
the GUP with the corresponding modified Stefan-Boltzmann law and the black hole temperature.
It turns out that black hole complementarity is still valid with the GUP for $N=1$.
Furthermore, in Sec.~\ref{sec:GUPMDR},
the duplication of information with the GUP can be evaded
in the framework of Gravity's Rainbow.
Finally, conclusion and discussion will be given in Sec.~\ref{sec:con}.


\section{Black hole complementarity}
\label{sec:Sch}

Let us encapsulate the gedanken experiment in the
Schwarzschild black hole by assuming the Heisenberg uncertainty principle of
$\Delta x \Delta p \ge 1$ and the ordinary dispersion relation for massless particles, $E^2-p^2=0$
\cite{Susskind:1993mu}.
In the Kruskal-Szekeres coordinates,
the metric of the Schwarzschild black hole is given by
\begin{equation}\label{eq:Kruskal1}
ds^2=-\frac{32 G^3 M^3}{r} e^{-\frac{r}{2GM}}dUdV ,
\end{equation}
where $U=\pm e^{-\frac{(t-r^*)}{4GM}}$,
$V=e^{\frac{(t+r^*)}{4GM}}$, and $r^*=r+2G M \ln \left(|r-2GM|/2GM\right)$.
The plus and minus signs in $U$ coordinate are for the inside and outside of the horizon,
respectively.
To obtain the Page time for the old black hole  \cite{Page:1993wv},
we consider the Stefan-Boltzmann law,
\begin{equation}\label{eq:SB1}
\frac{d M}{dt}=-A \sigma T^4,
\end{equation}
where $\sigma$ denotes the Stefan-Boltzmann constant, and
$A$ and $ T$ are the area and temperature of the black hole identified with $A=16 \pi G^2 M^2$ and $T=1/(8\pi G M)$
for the Schwarzschild black hole, respectively.
Then, the Page time $t_{\textrm{\tiny P}}$
can be calculated from the Stefan-Boltzmann law \eqref{eq:SB1} as
\begin{equation}\label{eq:pageT}
t_{\textrm{\tiny P}} \sim G^2 M^3,
\end{equation}
when the initial Bekenstein-Hawking entropy shrinks in half.

\begin{figure}[hpt]
  \begin{center}
  \includegraphics[width=0.5\linewidth]{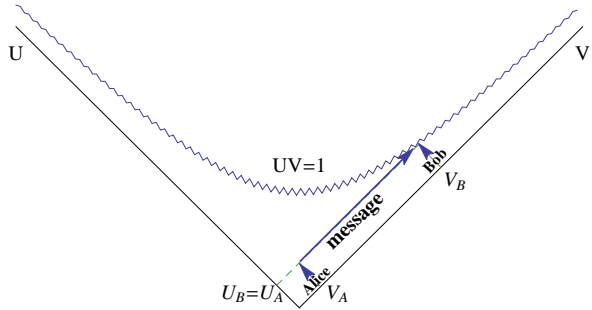}
  \end{center}
  \caption{For the Schwarzschild black hole (later the rainbow Schwarzschild black hole), the wiggly curve in $UV=1$ means the curvature singularity at the origin, $r=0$.
 Alice passes through the horizon at $V_{\rm A}$, and then,  after the Page time,
 Bob will jump into the horizon at $V_{\rm B}$.
 Alice should send the message with information to Bob at least at $U_{\rm A}$ before Bob hits the singularity.
 }
  \label{fig:penrose1}
\end{figure}


Now, we suppose that Alice first jumps into the horizon at $V_{\rm A}$,
and then Bob passes through the horizon with a record of his measurements of
information from the Hawking radiation after the Page time $t_{\textrm{\tiny P}}$  at $V_{\rm B}$
as shown in Fig. \ref{fig:penrose1}.
Alice should send the message with information to Bob
before he hits the curvature singularity,
$U_{\rm A}=U_{\rm B}=V_{\rm B}^{-1}= e^{-t_{\textrm{\tiny P}}/(4GM)} \sim e^{-G M^2}$.
So, the proper time $\Delta \tau$ for Alice to send the message to Bob at least at $U_{\rm A}$
can be calculated from the metric \eqref{eq:Kruskal1} near the horizon $r= r_{\rm H}$ as
$\Delta \tau^2 
\sim G^2 M^2 e^{-GM^2}, $
where $\Delta V_{\rm A}$ is a nonvanishing finite value near $V_{\rm A}$ for the free-fall \cite{Susskind:1993mu}.
Then, the energy-time uncertainty principle of  $\Delta E \Delta \tau \ge 1$
gives the required energy $\Delta E$
as
\begin{equation}\label{eq:DE}
\Delta E \sim \frac{1}{GM} e^{GM^2},
\end{equation}
which is definitely larger than the black hole mass, {\it i.e.}, $\Delta E \gg M$,
so that information must be encoded into the message with super-Planckian frequency.
Therefore, the duplication of information is impossible and black hole complementarity
can be well-defined.

\section{black hole complementarity with GUP}
\label{sec:GUPDR}

To find out the validity of the no-cloning theorem,
we calculate the required energy to duplicate information by employing
the modified temperature of the Schwarzschild black hole and
 the modified Stefan-Boltzmann law
which are commensurate with the GUP  \cite{AmelinoCamelia:2005ik}.
Let us start with the GUP defined by
 \cite{Maggiore:1993rv,  Maggiore:1993zu, Scardigli:1999jh, Adler:1999bu,  Hossenfelder:2012jw, Veneziano:1986zf, Gross:1987ar, Amati:1988tn, Konishi:1989wk},
\begin{equation}\label{eq:GUP}
\Delta x\Delta p\geq1+\alpha_{\textrm{\tiny GUP}} L_{\rm p}^{2}\Delta p^2,
\end{equation}
where  $\alpha_{\textrm{\tiny GUP}}$ is the GUP parameter
and the Planck length is denoted by  $L_{\rm p}=\sqrt G$.
The first modification is that the black hole temperature
from the GUP \eqref{eq:GUP} is given as~\cite{Adler:2001vs},
\begin{align}
T&=\frac{GM}{4\pi  \alpha_{\textrm{\tiny GUP}}L_{\rm p}^2}
\left(1-\sqrt{1-\frac{\alpha_{\textrm{\tiny GUP}}L_{\rm p}^2}{G^2M^2}}\right) \notag \\
&\simeq \frac{1}{8\pi GM}+\frac{\alpha_{\textrm{\tiny GUP}}}{32\pi G^2 M^3}. \label{eq:TGUP}
\end{align}

Next, let us derive the Stefan-Boltzmann law consistent  with the GUP
along the line of Ref. \cite{AmelinoCamelia:2005ik} in order to get the  Page time corrected by the GUP.
The wave lengths of photons in a cubical box with edges of length $L$
are subject to the boundary condition $1/\lambda=n/(2L)$ with a positive integer $n$.
For oscillators in the box, the energy density is written in an integral form
as
\begin{align}
\rho&=\frac{1}{V}\int\bar E~ g(\nu) \textrm d \nu \notag \\
&=2\int\bar E~  \textrm d^3 \nu, \label{eq:energy_density}
\end{align}
where $g(\nu)  d \nu$ is the number of modes
in an infinitesimal frequency interval $[\nu,\nu+\textrm d \nu]$
and $\bar E$ means the average energy  per oscillator given by
\begin{equation}\label{eq:Boltzmann}
\bar E=\frac{E}{e^{\frac E T}-1}.
\end{equation}
The relation for GUP \eqref{eq:GUP} should be reflected in the modification of the de Broglie relation as \cite{AmelinoCamelia:2005ik}
\begin{equation}\label{eq:de_Brogile_GUP}
\lambda = \frac 1 p \left(1+\alpha_{\textrm{\tiny GUP}}L_{\rm p}^2 p^2\right),
\end{equation}
and then one can read off the relation between the energy and the frequency by using the conventional
dispersion relation,
$\nu = E\left(1-\alpha_{\textrm{\tiny GUP}}L_{\rm p}^2 E^2 +\mathcal O (L_{\rm p}^4 E^4) \right)$.
Thus the energy density at a given temperature $T$ is calculated from Eq. \eqref{eq:energy_density} as
\begin{align}
\rho&=8\pi\int\textrm d E\frac{E^3}{e^{\frac E T}-1}
\left(1-5\alpha_{\textrm{\tiny GUP}}L_{\rm p}^2 E^2+\mathcal O (L_{\rm p}^4E^4)\right) \label{eq:integral} \\
&\simeq8\pi T^4\int\textrm d \xi \frac{\xi^3}{e^{\xi}-1}- 40\pi\alpha_{\textrm{\tiny GUP}}L_{\rm p}^2 T^6 \int\textrm d \xi \frac{\xi^5}{e^{\xi}-1}
\label{eq:integral2}
\\
&\simeq \frac{8\pi^5}{15}T^4-\frac{320\pi^7}{63}\alpha_{\textrm{\tiny GUP}}L_{\rm p}^2T^6, \label{eq:integral3}  
\end{align}
where $ \xi =E/ T$ at the finite temperature $T$.

It is worth noting that
the  expression \eqref{eq:integral} is actually
valid as long as $E < E_{\rm M} = (\sqrt{\alpha_{\textrm{\tiny GUP}}} L_{\rm p})^{-1}$,
where the integral involved in Eq.~\eqref{eq:integral} should be integrated up to
the finite value of $E_{\rm M}$ rather than infinity due to the constraint of the GUP \eqref{eq:GUP}.
In fact, the values of the two respective integrals in Eq.~\eqref{eq:integral2}
 up to the cutoff of $\xi_{\rm M}=(\sqrt{\alpha_{\textrm{\tiny GUP}}} L_{\rm p} T)^{-1}$
  are expected to be slightly less than those values integrated up to infinity
  since the integrands are positive definite.
 For the sake of our neat calculation, we release the upper bound up to infinity,
 and then obtain the larger exact coefficients for each power of the temperature in Eq.~\eqref{eq:integral3}. However, for the large black hole, these coefficients
 are insensitive to the final results.
 This kind of approximation will also be used in the later calculations.

Next, the Stefan-Boltzmann law improved by the GUP for the evaporating black hole is
obtained as
\begin{equation}\label{eq:SBlawGUP}
\frac{\textrm d M}{\textrm d t} \simeq -A\left(\frac{8\pi^5}{15}T^4-\frac{320\pi^7}{63}\alpha_{\textrm{\tiny GUP}}L_{\rm p}^2T^6\right)
\end{equation}
up to the linear order of $\alpha_{\textrm{\tiny GUP}}$,
where $A$ denotes the area of the black hole.
The entropy can also be calculated by use of
the first law of black hole thermodynamics as $S =\int 1/T \textrm d M =4\pi GM^2-2\pi\alpha_{\textrm{\tiny GUP}}\textrm{ln} (\sqrt G M)$
 \cite{AmelinoCamelia:2005ik}.
\begin{figure}[hpt]
  \begin{center}
  \includegraphics[width=0.6\linewidth]{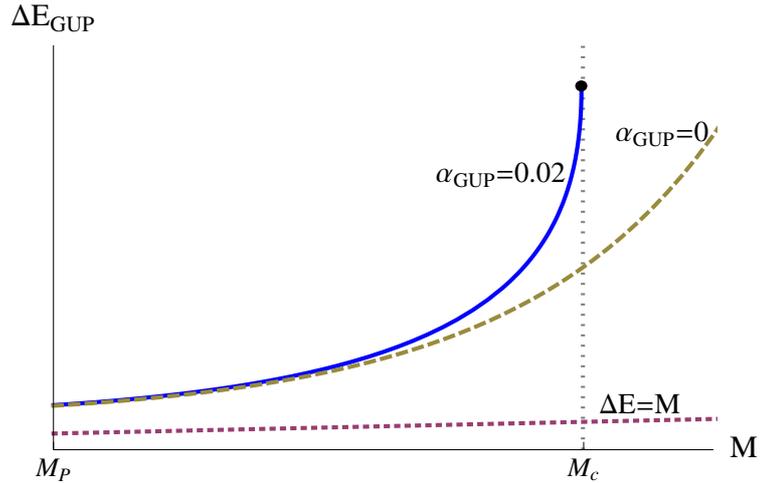}
  \end{center}
  \caption{ 
The standard result \eqref{eq:DE} with Heisenberg uncertainty principle
   ($\alpha_\textrm{\tiny GUP}=0$) and Eq.~\eqref{eq:DEGUP} for the GUP ($\alpha_\textrm{\tiny GUP}=0.02$)
    are plotted, respectively. There are upper bounds at $M_{\rm c}$ for the case of GUP.
As the black hole mass $M$ increases, the energy denoted by $\Delta E=M$ linearly increases with a very small slope depending on our scale while the required energy for the GUP increases exponentially.
 So, the required energy $\Delta E_{\textrm{\tiny GUP}}$ is always
     larger than the black hole mass represented by the dotted line of $\Delta E_{\rm GUP}=M$.
  The Planck mass is chosen as $M_{\rm P}=1/L_{\rm P}=1$ for simplicity.}
  \label{fig:GUP}
\end{figure}

Using the temperature \eqref{eq:TGUP} and the modified Stefan-Boltzmann law \eqref{eq:SBlawGUP},
one can get the Page time
when the black hole has emitted half of its initial entropy and the information of the black hole starts to be emitted by the Hawking radiation
as
\begin{align}
t_{\textrm{\tiny P}}(M)
=\int^{M}_{M_{\rm Page}} \textrm d M \frac{1}{16\pi G^2 M^2 \left(\frac{8\pi^5}{15}T^4-\frac{320\pi^7}{63}\alpha_{\textrm{\tiny GUP}}L_{\rm p}^2T^6\right)},
\end{align}
where $M_{\rm Page}$ is the mass of the black hole at the Page time, which is smaller than the initial mass  $M$ as $M_{\rm Page} \ll M$.
Following the argument of Ref.~\cite{Page:1993wv},
we rewrite the page time in terms of the black hole mass up to the subdominant term for the large black hole as
\begin{equation}\label{eq:}
t_{\textrm{\tiny P}}(M) \sim G^2 M^3-\alpha_{\textrm{\tiny GUP}} G M.
\end{equation}

From the Schwarzschild metric \eqref{eq:Kruskal1},
the interval of the proper time $\Delta \tau$ which is nothing but the free-fall time for Alice near the horizon of $r = 2GM$
 is given as
$\Delta\tau\sim GMe^{-t_{\textrm{\tiny P}}(GM)^{-1}}$.
One can find the appropriate energy-time uncertainty principle
as $\Delta \tau \Delta E \ge 1+\alpha_{\textrm{\tiny GUP}} L_{\rm p}^{2}\Delta E^2$
from the GUP  \eqref{eq:GUP}.
Finally,
the required energy is read off
 from the generalized energy-time uncertainty principle as
\begin{align}
\Delta E_{\textrm{\tiny GUP}} \sim
\frac{M}{2\alpha_{\textrm{\tiny GUP}}}e^{-GM^2+ \alpha_{{\tiny GUP}}}
\left(1-\sqrt{1-\frac{4\alpha_{\textrm{\tiny GUP}}}{GM^2}e^{2(GM^2-\alpha_{\textrm{\tiny GUP}})}}\right), \label{eq:DEGUP}
\end{align}
where it nicely reduces to Eq. \eqref{eq:DE} for $\alpha_{\textrm{\tiny GUP}} \to 0$.

Even though the required energy \eqref{eq:DEGUP} has an upper bound $M_{\rm c}$,
 the energy is larger than the mass of the black hole as shown in Fig.~\ref{fig:GUP}.
It shows that the GUP effect improves the no-cloning theorem in the sense that
the required energy for a given black hole mass
 is larger than that without the GUP correction, {\it i.e.},
$\Delta E_{\textrm{\tiny GUP}} \ge \Delta E \gg M$, so that
black hole complementarity is still valid even for $N=1$
when the appropriate temperature and the Stefan-Boltzmann law are
employed.
However, this is not the whole story since this approach is incomplete in the sense that the minimal length should be treated as the invariant scale, so that the issue should be discussed in the regime of Gravity's Rainbow.


\section{black hole complementarity with GUP in Gravity's Rainbow}
\label{sec:GUPMDR}

In order to avoid the length contraction of the minimal length due to the FitzGerald-Lorentz contraction in Einstein's relativity theory, we introduce the MDR in the doubly special relativity \cite{AmelinoCamelia:2000ge, AmelinoCamelia:2000mn} which makes the Planck length invariant as a minimal length.
Under the deformed symmetries, the measure of integration in momentum space should be modified in order for the relativistic properties not to be spoiled.
However, in our case, it will turn out that the measure is invariant.

By using the non-linear Lorentz transformation in the momentum space, the MDR can be compactly written as \cite{Magueijo:2001cr, Magueijo:2002am}
%
%
\begin{equation}\label{MDR1}
f(E)^2 E^2-g(E)^2 p^2 = m^2,
\end{equation}
where
the rainbow functions $f(E)$ and $g(E)$ satisfy  $\lim_{E\rightarrow 0} f =1$ and $\lim_{E\rightarrow 0} g =1$,
and
$E$ and $m$ denote the energy and the mass of the test particle,
respectively.
The metric tensor associated with the MDR \eqref{MDR1}
is expressed in terms of a one-parameter family of orthonormal frame fields
based on the modified equivalence principle as
$g^{\mu\nu}(E)= \eta^{ab} e_a^\mu (E) e_b^\nu (E),$
where $e_0(E)=f^{-1}(E)\tilde{e}_0$ and $e_i(E)=g^{-1}(E)\tilde{e}_i$,
and $\tilde{e}$ is the ordinary energy-independent vielbein.
Then, the energy-dependent Schwarzschild metric is obtained as
 \cite{Magueijo:2002xx}
\begin{align}\label{eq:rainbowmetric}
\textrm d s^2=-\frac{1}{f(E)^2}\left(1-\frac{2GM}{r}\right)\textrm d t^2+\frac{1}{g(E)^2}\frac{1}{\left(1-\frac{2GM}{r}\right)}\textrm d r^2+\frac{r^2}{g(E)^2}\textrm d \Omega^2,
\end{align}
which is called the rainbow Schwarzschild black hole.
From now on,
we will employ the rainbow functions proposed in Ref.~\cite{AmelinoCamelia:2005ik},
\begin{equation}\label{fandg}
f(E)=\left(1+\frac{\beta_{\textrm{\tiny MDR}}}{2}L_{\rm p} E+\left(\frac 1 2\gamma_{\textrm{\tiny MDR}}-\frac 1 8 \beta_{\textrm{\tiny MDR}}\right)L_{\rm p}^2 E^2\right),
 \qquad  g(E)=1,
\end{equation}
where $\beta_{\textrm{\tiny MDR}}$ and $\gamma_{\textrm{\tiny MDR}}$ are the MDR parameters.
Then, the MDR \eqref{MDR1} is rewritten for a massless particle as
\begin{equation}\label{eq:MDR}
p = E\left(1+\frac{\beta_{\textrm{\tiny MDR}}}{2}L_{\rm p} E+\left(\frac 1 2\gamma_{\textrm{\tiny MDR}}-\frac 1 8 \beta_{\textrm{\tiny MDR}}\right)L_{\rm p}^2 E^2\right).
\end{equation}

Now, we are in a position to derive the black hole temperature and
the modified Stefan-Boltzmann law by considering
not only the GUP \eqref{eq:GUP} but also the MDR \eqref{eq:MDR}.
Combining the modified de Broglie relation \eqref{eq:de_Brogile_GUP} and the MDR \eqref{eq:MDR}
 gives the relation between the energy $E$ and the frequency $\nu$,
\begin{align}
\nu = E\left(
1+\frac 1 2 \beta_{\textrm{\tiny MDR}}L_{\rm p}E+\left(
\frac 1 2 \gamma_{\textrm{\tiny MDR}}-\frac 1 8 \beta_{\textrm{\tiny MDR}}^2-\alpha_{\textrm{\tiny{GUP}}}
\right)L_{\rm p}^2 E^2 \right) +\mathcal O (L_{\rm p}^3 E^3)
\end{align}
with the assumptions of $ \Delta p =  p$  and $ \Delta E = E$ since the momentum and the energy uncertainties will be of order of the momentum $p$ and the energy $E$, respectively \cite{Gim:2015zra}.

The measure of the energy density \eqref{eq:energy_density}  is rewritten
by the momentum $p$ in terms of 
the de Broglie relation \eqref{eq:de_Brogile_GUP} as
\begin{align}
\rho
=2 \int \bar E \textrm d^3 \left(  \frac{ g(E)p}{ \left(1+\alpha_{\textrm{\tiny GUP}}L_{\rm p}^2 g(E)^2 p^2\right)}\right), \label{eq:momentum_int}
\end{align}
where 
the measure of integration in momentum space should be modified under  the deformed symmetries, for example, in the context of modified thermodynamics in Ref.~\cite{Gorji:2016gfr}.
However, the measure of  Eq.~\eqref{eq:momentum_int} includes the three-dimensional momentum $p$,
so that the measure of integration  for  the deformed symmetries is invariant
 since  the rainbow function corresponding to the momentum is $g(E)=1$.

Then, the energy density \eqref{eq:momentum_int} is calculated
with the average energy \eqref{eq:Boltzmann} per oscillator and the MDR\eqref{eq:MDR},
\begin{align}
\rho
&=8\pi\int\textrm d E\frac{E^3}{e^{\frac E T}-1}
\left(
1+2\beta_{\textrm{\tiny MDR}}L_{\rm p}E+\left(
\frac 5 2 \gamma_{\textrm{\tiny MDR}}+\frac 5 8 \beta_{\textrm{\tiny MDR}}^2-5\alpha_{\textrm{\tiny GUP}}
\right)L_{\rm p}^2 E^2+\mathcal O (L_{\rm p}^3 E^3)
\right) \notag \\
&\simeq \frac{8\pi^5}{15}T^4+384\pi\zeta(5)\beta_{\textrm{\tiny MDR}}L_{\rm p}T^5+\left(
\frac 1 2 \gamma_{\textrm{\tiny MDR}}+\frac 1 8 \beta_{\textrm{\tiny MDR}}^2-\alpha_{\textrm{\tiny GUP}}
\right)\frac{320\pi^7}{63} L_{\rm p}^2 T^6, \label{eq:19}
\end{align}
where the similar approximations to Eq.~\eqref{eq:integral3} are used,
so that the modified Stefan-Boltzmann law induced by the GUP and the MDR is given as
\begin{equation}\label{eq:SBlawGUPMDR}
\frac{\textrm d M}{\textrm d t}
\simeq -A\left(\frac{8\pi^5}{15}T^4+384\pi\zeta(5)\beta_{\textrm{\tiny MDR}}L_{\rm p}T^5+\left(
\frac 1 2 \gamma_{\textrm{\tiny MDR}}+\frac 1 8 \beta_{\textrm{\tiny MDR}}^2-\alpha_{\textrm{\tiny GUP}}
\right)\frac{320\pi^7}{63} L_{\rm p}^2 T^6\right),
\end{equation}
where $A$ is the area of the black hole.

Next, plugging the MDR \eqref{eq:MDR} into the GUP \eqref{eq:GUP},
one can get the following relation,
 \begin{align}
E \left(1+ \beta_{\textrm{\tiny MDR}} L_{\rm p} E + \left(\frac{1}{2}\gamma_{\textrm{\tiny MDR}}-\frac{1}{8}\beta_{\textrm{\tiny MDR}}\right) L_{\rm p}^2 E^2 \right) \Delta x   \simeq 1+\alpha_{\textrm{\tiny GUP}}L_{\rm p}^2 E^2 + \mathcal{O}(L_{\rm p}^3 E^3) \label{eq:TT}
\end{align}
 by assuming that $ \Delta p =  p$ since the momentum is of order of the momentum $p$ \cite{Adler:2001vs, Gim:2015zra}.
Next, by use of  $\Delta x = 2GM$ and $E=4\pi T$ \cite{Adler:2001vs},
the black hole temperature associated with the GUP with the MDR
is obtained as
 \begin{equation}\label{eq:TMDRGUP}
T \left( 1+ \beta_{\textrm{\tiny MDR}} L_{\rm p} (4\pi T) + \left(\frac{1}{2}\gamma_{\textrm{\tiny MDR}}-\frac{1}{8}\beta_{\textrm{\tiny MDR}}\right) L_{\rm p}^2 (4\pi T)^2 \right)  \simeq \frac{1}{8\pi GM} \left( 1+\alpha_{\textrm{\tiny GUP}}L_{\rm p}^2 (4\pi T)^2 \right),
\end{equation}
where we neglected the higher-order correction terms above the square of $L_{\rm p}^2 T^2$ in Eq.~\eqref{eq:TT} since the Planck length and the black hole temperature of the large black hole are very small.
And the black hole temperature \eqref{eq:TMDRGUP} gives rise to the entropy of the black hole as
 $S \sim A/4 + \beta_{\textrm{\tiny MDR}} \sqrt{A} + (\alpha_{\textrm{\tiny GUP}} - \gamma_{\textrm{\tiny MDR}}/2)\ln A $.

 Note that the expression \eqref{eq:TMDRGUP} is a non-linear closed form,
  so that we will choose one of the simplest but non-trivial combination of parameters,
\begin{equation}\label{eq:parameter}
\beta_{\textrm{\tiny MDR}}=0,\qquad  \alpha_{\textrm{\tiny GUP}} = \frac 1 2 \gamma_{\textrm{\tiny MDR}},
\end{equation}
which makes the entropy of the black hole to be the Bekenstein-Hawking entropy.
For the special choice of parameters
\eqref{eq:parameter},
the rainbow functions
\eqref{fandg} are rewritten as
$f(E)=1+\alpha_{\textrm{\tiny GUP}} L_{\rm p}^2 E^2$ and $g(E)=1$.
And the modified Stefan-Boltzmann law \eqref{eq:SBlawGUPMDR} and the temperature \eqref{eq:TMDRGUP}
are simplified as
$\textrm d M/\textrm d t \simeq -A (8\pi^5/15)T^4$ and $T\simeq (8\pi GM)^{-1}$.
They are nothing but the Hawking temperature and the conventional Stefan-Boltzmann law,
so that the Page time is simply written as $t_{\textrm{\tiny P}}\sim G^2M^3$.

Now, let us calculate the required energy for the duplication of information in the rainbow Schwarzschild black hole
along the argument in Ref.~\cite{Susskind:1993mu}.
 First, the rainbow metric \eqref{eq:rainbowmetric} is written
 in terms of the rainbow Kruskal-Szekeres coordinates defined as \cite{Gim:2015zra}
\begin{align}
\textrm d s^2 =-\frac{4r_{\rm H}^3}{g(E)^2 r}e^{-\frac{r}{r_{\rm H}}}\textrm d U\text d V, \label{eq:kruskalMDR}
\end{align}
where $U=\pm e^{-\left((g/f)t-r^{*}\right)/(2GM)}$,
$V=e^{\left((g/f)t+r^{*}\right)/(2GM)}$
and $r^*= r + 2G M \ln \left(|r-2GM|/2GM\right)$.
The plus sign in $U$ coordinate will be selected to describe the inside of the horizon.
As shown in Fig.~\ref{fig:penrose1}, Alice should send her information encoded into a message before
 $U_{\rm A}=U_{\rm B}=V_{\rm B}^{-1} \sim e^{-(g/f)G M^2}$,
so that
the proper time measured by Alice near the horizon $r = 2GM$ is obtained
from the metric \eqref{eq:kruskalMDR} as
\begin{equation}\label{eq:Dtau2}
\Delta\tau \sim GM e^{-\frac{GM^2}{1+\alpha_{\textrm{\tiny GUP}}  L_{\rm p}^2 \Delta E^2}},
\end{equation}
where we assumed that $\Delta V_{\rm A}$ is a nonvanishing finite value \cite{Susskind:1993mu}.

\begin{figure}[hpt]
  \begin{center}
  \includegraphics[width=0.6\linewidth]{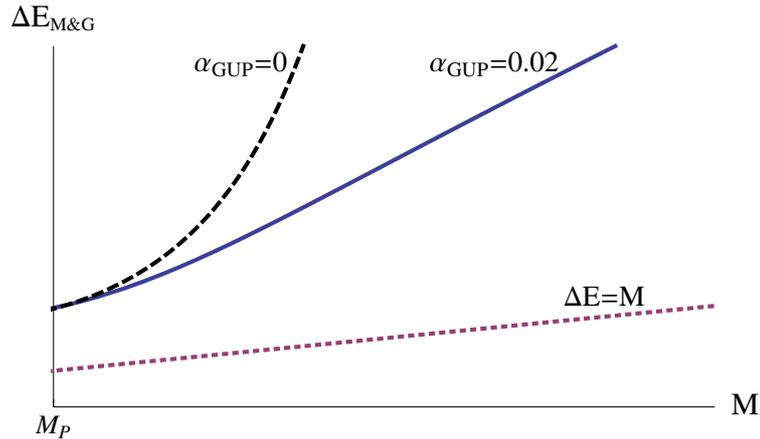}
  \end{center}
  \caption{We set the Planck length to $L_{\rm p}=1$.
   The upper thick dotted curve is for $\alpha_{\textrm{\tiny GUP}} =0$ corresponding to
    the standard result using the Heisenberg uncertainty principle and the ordinary dispersion relation, whereas the solid curve is for $\alpha_{\textrm{\tiny GUP}} =0.02$
    corresponding to the required energy corrected by the GUP with the MDR.
As the black hole mass $M$ increases, the energy described by $\Delta E=M$ linearly increases with a very small slope while the required energy corrected by the GUP and the MDR increases exponentially.
So, the required energy $\Delta E_{\textrm{\tiny M\&G}}$ is larger than the mass of the black hole.
    Note that we cut off the other branch of the Lambert W-function since it describes the small mass region below the Planck mass $M_{\rm P}$.
}
  \label{fig:GUPMDR}
\end{figure}

Next, one can find
the appropriate energy-time uncertainty principle from the GUP and the MDR
as
$\Delta\tau\Delta E \ge 1+\alpha_{\textrm{\tiny GUP}}  L_{\rm p}^2\Delta E^2+\mathcal O ( \alpha_{\textrm{\tiny GUP}} ^2 L_{\rm p}^4 \Delta E^4)$
by use of the definition of the group velocity as $v_{\rm G}= \Delta E/\Delta p$.
Then,
the required energy $\Delta E_\textrm{\tiny M\&G}$ for duplication of information
is finally obtained as
\begin{equation}\label{eq:DEGUPMDR}
\Delta E_\textrm{\tiny M\&G} \sim \frac{1+\alpha_{\textrm{\tiny GUP}}  L_{\rm p}^2\Delta E_\textrm{\tiny M\&G}^2}{GM} e^{\frac{GM^2}{1+\alpha_{\textrm{\tiny GUP}}  L_{\rm p}^2 \Delta E_\textrm{\tiny M\&G}^2}},
\end{equation}
which goes to the ordinary relation
for $\alpha_{\textrm{\tiny GUP}} \to 0$.
Since it is non-trivial
to solve Eq.~\eqref{eq:DEGUPMDR} with respect to $\Delta E_\textrm{\tiny M\&G}$,
and so we solve it for $M$ as
$M=\sqrt{
-(L_{\rm p}^{-2}/2)(1+\alpha_{\textrm{\tiny GUP}}  L_{\rm p}^2 \Delta E_\textrm{\tiny M\&G}^2 ) W\left(Y(\Delta E_\textrm{\tiny M\&G})\right)
},$
 where  the Lambert W-function is defined as $Y=W(Y)e^{W(Y)}$ with the variable $Y = -2 (1+\alpha_{\textrm{\tiny GUP}}  L_{\rm p}^2 \Delta E_\textrm{\tiny M\&G}^2) \Delta E_\textrm{\tiny M\&G}^{-2}L_{\rm p}^{-2}$.
Then,
we can demonstrate the behavior of $\Delta E_\textrm{\tiny M\&G}$ with respect to $M$ by a parametric plot of a curve for the points  $(M,\Delta E_\textrm{\tiny M\&G})$   in Fig.~\ref{fig:GUPMDR}.
The required energy $\Delta E_\textrm{\tiny M\&G}$ to send the message from Alice to Bob before he hits the singularity always exceeds the black hole mass $M$.
Thus it indicates that
the no-cloning theorem in quantum theory for black hole complementarity can be made valid
in the extended regime of the GUP in Gravity's Rainbow.

\section{Conclusion and discussion}
\label{sec:con}

The required energy for Alice to send the message to Bob who jumped into the black hole at the Page time was calculated in the presence of the minimal length defined by the GUP.
We showed that the required energy becomes the super-Planckian scale,
so that it turned out that the unitarity in quantum mechanics is maintained and black hole complementarity is safe.
Furthermore,
we revisited the above issue by employing MDR
for the rainbow Schwarzschild black hole.
The required energy also exceeds the mass of black hole,
so that the no-cloning theorem for black hole complementarity can be made valid.
Unfortunately, it is not a general proof since the present calculation is based on the particular rainbow functions and the special choice of parameters.

So, we mention the reason why the specific combination \eqref{eq:parameter}
of the GUP and MDR parameters was adopted.
The parameter $\beta_{\textrm{\tiny MDR}}$ in the MDR \eqref{eq:MDR}
gives rise to
the square root of the black hole area $A$
as the leading correction of the black hole entropy of
 $S \sim A/4 + \beta_{\textrm{\tiny MDR}} \sqrt{A} + \mathcal{O}(\ln A) $  \cite{AmelinoCamelia:2005ik}.
But it has been shown that a logarithmic correction 
appears as the leading correction of the entropy-area
relation  in various quantum gravity scenarios \cite{Solodukhin:1997yy, Carlip:2000nv, Meissner:2004ju,  Arzano:2005jt}.
Note that in loop quantum gravity,
such a $\sqrt{A}$ correction to the entropy of the black hole was
excluded \cite{Rovelli:1996dv, Ashtekar:1997yu, Kaul:2000kf}, so that
 the presence of linear-in-$L_{\rm p}$ contributions related to  $\beta_{\textrm{\tiny MDR}}$  in the MDR was also eliminated \cite{AmelinoCamelia:2004xx, AmelinoCamelia:2005ik}.
In these respects,
we took the parameter $\beta_{\textrm{\tiny MDR}}$ to vanish in our work.
Finally, we obtained our results by assuming the additional choice of parameters of $\alpha_{\textrm{\tiny GUP}} =  \gamma_{\textrm{\tiny MDR}}/2$
where the entropy of the black hole respects the area law eventually.
We hope that the present study will be extended to generic models in the near future.

Finally, we would like to discuss whether the required energies \eqref{eq:DEGUP}
and \eqref{eq:DEGUPMDR}
are still larger than the mass of the black hole when the Hawking radiation consists of a large number of scalar fields.
Our results can be extended to the case that
the black hole emits the large number $N$ of species of massless scalar fields
under the large $N$ rescaling scheme as
$M \rightarrow M' \equiv \sqrt{N}M$ along the lines of Ref.~\cite{Chen:2014bva}.
In this scheme, all plots in Fig.~\ref{fig:GUP} and Fig.~\ref{fig:GUPMDR} are just rescaled along  the $M$-axis without
changing any physical deformation,
so that we can show that the required energies \eqref{eq:DEGUP} and \eqref{eq:DEGUPMDR} with a large $N$ still exceed
the mass of the black hole.
Consequently, despite the large number of scalar fields,
it turns out that the violation of black hole complementarity can be evaded.

\acknowledgments

This work was supported by
the National Research Foundation of Korea(NRF) grant funded by the
Korea government(MSIP) (2017R1A2B2006159).


\bibliographystyle{JHEP}       

\bibliography{references}

\end{document}